\documentclass[12pt]{article}
\usepackage[T2A]{fontenc}
\usepackage[utf8]{inputenc}
\usepackage[english]{babel}
\usepackage{amssymb}
\usepackage{amsmath}
\usepackage{graphicx}
\usepackage[small,sc,center]{titlesec}
\usepackage{indentfirst}
\usepackage{siunitx}
\usepackage[labelsep=period]{caption}

\newcommand{\nc}{\newcommand}
\nc{\mzams}{M_\mathrm{ZAMS}}
\nc{\tev}{t_\mathrm{ev}}
\begin{document}

\begin{center}
\textbf{Evolution and period change in RR~Lyr variables of the globular cluster M~3}

\vskip 3mm
\textbf{Yu. A. Fadeyev\footnote{E--mail: fadeyev@inasan.ru}}

\textit{Institute of Astronomy, Russian Academy of Sciences,
        Pyatnitskaya ul. 48, Moscow, 119017 Russia} \\

Received May 15, 2018
\end{center}

\textbf{Abstract} ---
The grid of evolutionary tracks of population II stars with initial masses
$0.81M_\odot\le\mzams\le0.85M_\odot$ and chemical composition of the globular
cluster M~3 is computed.
Selected models of horizontal branch stars were used as initial conditions for
solution of the equations of radiation hydrodynamics and time--dependent
convection describing radial stellar oscillations.
The boundaries of the instability strip on the Herztsprung--Russel diagram
were determined using $\approx 100$ hydrodynamic models of RR~Lyr
pulsating variables.
For each evolutionary track crossing the instability strip the pulsation
period was determined as a function of evolutinary time.
The rate of period change of most variables is shown to range within
$-0.02\le\dot\Pi\le 0.05~\mbox{day}/10^6\mbox{yr}$.
Theoretical estimate of the mean period change rate obtained by the population
synthesis method is $\langle\dot\Pi\rangle=6.0\times 10^{-3}~\mbox{day}/10^6\mbox{yr}$
and agrees well with observations of RR~Lyr variables of the globular cluster M~3.
.

Keywords: \textit{stars: variable and peculiar}

\newpage
\section{introduction}

In each globular cluster the pulsating RR~Lyr type variables represent the
homogeneous group of stars with nearly the same age and small spread in their
abundances, so that they are the useful tool for investigation of the stellar cluster.
RR~Lyr variables are thought to be the stars with masses of $M\approx 0.6M_\odot$
at the core helium--burning stage of their evolution (Caputo 1998).
In contrast to more massive stars (i.e. with masses $M\gtrsim 2M_\odot$)
the onset of helium burning occurs in conditions of strong degeneracy of
the electron gas and leads to explosive energy generation growth.
A rapid rise (on the time scale $\lesssim 10^2$~yr) of energy
generation in the helium core exceeds the stellar luminosity
by six or seven orders of magnitude and is responsible for the
short--term cessation of hydrogen buring in the outer shell.
The star leaves the tip of the red giant branch and its luminosity
decreases by as much as two orders of magnitude.
Since the pioneering work of Schwarzschild and H\"arm (1962) the core helium flash
was studied in detail by many authors but its short time scale was a major obstacle
in observational detection and corroboration of the theory.
Thus, comparison of evolution and pulsation models of low--mass population II
stars with observations of RR~Lyr type variables is of great interest because
it allows us to demonstrate the validity of the basic theoretical conclusions
on core helium flash and stellar evolution in globular clusters.

M~3 (NGC~5272) is one of the most studied globular clusters with plenty
of RR~Lyr variables.
$O-C$ diagrams constructed for 129 variables observed during the last 120 yr
allowed a determination of their secular period change rates
(Jurcsik et al. 2012).
In accord with stellar evolution theory the periods of RR~Lyr variables
decrease or increase.
However for comparison of the theory with observations of most interest
is the observational estimate of the mean  period change rate
$\dot\Pi\approx 0.01~\mbox{d}/10^6\mbox{yr}$ (Jurcsik et al. 2012).
Unfortunately, theoretical estimates of period change rates in RR~Lyr stars
obtained by Sweigart and Renzini (1979), Lee (1991) and Koopman et al. (1994)
cannot be compared with observations in detail since they were obtained
without application of the stellar pulsation theory.

The goal of this study is to fill this gap by obtaining estimates of period
change rates from consistent computations of the stellar evolution and
nonlinear stellar pulsations.
Initial fractional mass abundances of helium and heavier elements
($Y=0.25$, $Z=0.001$)
used for evolutionary computations correspond to the composition of
RR~Lyr stars in the globular cluster M~3
(Silva Aguirre et al. 2010; Denissenkov et al. 2017).

\section{evolution of horizontal branch stars}

Similar to our previous works on secular period changes in Cepheids
(Fadeyev 2013, 2014, 2015)
in this study we employed the method based on consistent computations
of stellar evolution and nonlinear stellar pulsations.
In particular, hydrodynamic computations of radial stellar oscillations are
carried out with initial conditions represented by selected stellar models
of evolutionary sequences.
Consistency between evolutionary and hydrodynamic models is also due to
the use of the same equation of state and opacity data.

Stellar evolutionary computations were carried out with the MESA code version 10108
(Paxton et al. 2011, 2013, 2015, 2018) from the zero age main sequence
up to exhaustion of helium in the core.
The ratio of the mixing length to the pressure scale height was assumed to be
$\alpha_\mathrm{MLT} =  2.0$.
To evaluate the mass loss rate due to the stellar wind we used the
Reimers (1975) formula
\begin{equation}
\label{freim}
\dot M_\mathrm{R} = 4\times 10^{-13} \eta_\mathrm{R} (L/L_\odot)(R/R_\odot) (M/M_\odot)^{-1} ,
\end{equation}
where $\eta_\mathrm{R}=0.5$.
On the whole we computed seven evolutionary sequences in the range
$0.81M_\odot\le\mzams\le 0.84M_\odot$ with step $\Delta\mzams=0.005M_\odot$
and one more evolutionary sequence with $\mzams=0.85M_\odot$.
The stellar mass just before the core helium flash
(i.e. at the tip of the red giant branch)
increases with $\mzams$ and for our evolutionary models ranges within
$0.60\le M\le 0.66M_\odot$.

\begin{figure}
\centerline{\includegraphics[width=15cm]{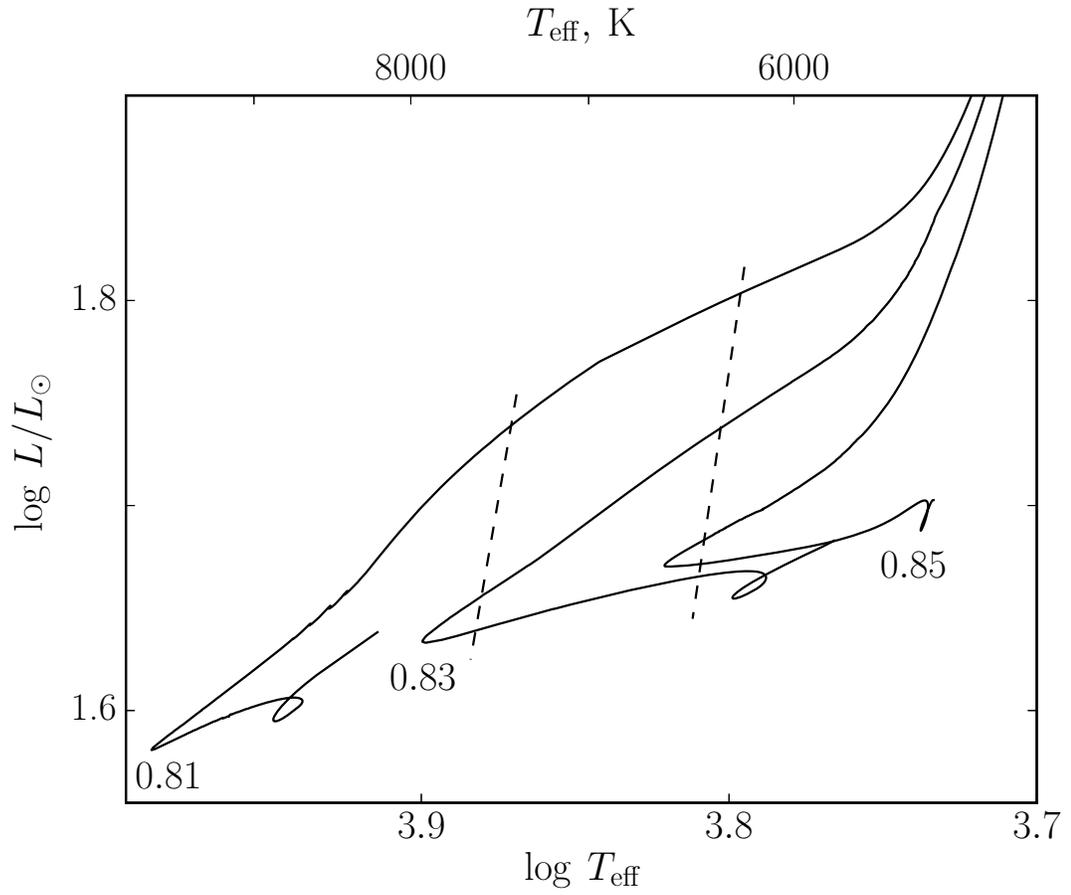}}
\caption{Evolutionary tracks of horizontal branch stars with initial masses
$\mzams=0.81M_\odot$, $0.83M_\odot$ and $0.85M_\odot$.
The instability strip boundaries are shown in dashed lines.}
\label{fig1}
\end{figure}

Fig.~\ref{fig1} shows parts of evolutionary tracks of stars with initial masses
$\mzams=0.81M_\odot$, $0.83M_\odot$ and $0.85M_\odot$
on the Hertzsprung--Russel diagram (HRD).
Each shown track begins at the point of the zero age horizontal branch (ZAHB)
corresponding to the onset of steady--state core helium burning.
The horizontal branch lifetime insignificantly increases with $\mzams$ and for
evolutionary sequences shown in Fig.~\ref{fig1} ranges from $1.05\times 10^8$
to $1.09\times 10^8$ years.
On the HRD evolution along the shown tracks proceeds clockwise.

Dashed lines in Fig.~\ref{fig1} show the instability strip boundaries
determined in the present study on the base of computed hydrodynamic models
(see below).
As clearly seen in Fig.~\ref{fig1}, the initial mass $\mzams=0.85M_\odot$ is near
the upper limit of RR~Lyr initial masses because evolutionary tracks with higher
$\mzams$ are beyond the red edge of the instability strip.
The star age at the ZAHB point increases with decreasing initial mass from
$10.8\times 10^9$ yr for $\mzams=0.85M_\odot$ to
$12.9\times 10^9$ yr for $\mzams=0.81M_\odot$,
whereas the recent estimates of the age of the globular cluster M~3
are $12.5\times 10^9$ yr (Dotter et al. 2011, VandenBerg et al. 2016)
and $12.6\times 10^9$ yr (Denissenkov et al., 2017).
Among evolutionary tracks computed in the present work the ZAHB stellar age
of the evolutionary sequence $\mzams=0.815M_\odot$ ($12.6\times 10^9$ yr)
most closely mathes these estimates.

\section{hydrodynamic models of RR~Lyr type stars}

Solution of the Cauchy problem for the equations of radiation hydrodynamics and
time--dependent convection was carried out in diffusion approximation for
radiation trasfer and trasport equations for turbulent convection
(Kuhfu\ss\ 1986).
The system of equations and main parameters of the problem are discussed
in one of our previous papers (Fadeyev 2015).
In the present work we computed nearly a hundred hydrodynamic models for
eight evolutionary tracks.

The period of radial oscillations was evaluated by the discrete Fourier transform
of the kinetic energy of pulsation motions
\begin{equation}
 E_\mathrm{K}(t) = \frac{1}{2}\sum\limits_{j=1}^N \Delta M_j U(t)_j^2
\end{equation}
on the time interval comprising several hundred pulsation cycles.
Here $N=400$  is the number of mass zones in the hydrodynamical model,
$\Delta M_j$ is the mass of the $j$--th zone,
$U(t)_j$ is the gas flow velocity.

For all hydrodynamic models together with the pulsation period $\Pi$ we evaluated
the growth rate $\eta=\Pi^{-1} d\ln E_\mathrm{K,max}/dt$
on the initial interval of integration of the equations of hydrodynamics.
An inverse of the growth rate is the number of periods corresponding to
change of $E_\mathrm{K,max}$ by a factor of $e=2.718\ldots$.
If the star is unstable against radial oscillations ($\eta>0$)
then the linear growth of $\ln E_\mathrm{K,max}$ is followed by the limit cycle
stage when the pulsation amplitude is time--independent.

Integration of the equations of hydrodynamics for stable models is accompanied
by nearly linear decrease in $\ln E_\mathrm{K,max}$.
The finite amplitude of initial motions of the stellar model stable against
radial oscillations is due to errors arising when the zonal quantities
are calculated by interpolation of the selected evolutionary model.
Solution of the equations of hydrodynamics for models with decaying
oscillations allowed us to evaluate the oscillation period $\Pi$ and
the growth rate ($\eta<0$).

The evolutionary time $\tev$ and pulsation period $\Pi$ at the boundary of the
instability strip ($\eta=0$) were obtained by interpolation between two
adjacent hydrodynamical models with opposite signs of $\eta$.
Location of the instability strip boundaries on the HRD determined for eight
evolutionary tracks is shown in Fig.~\ref{fig1}.

During evolution from one boundary of the instability strip to another
RR~Lyr type stars undergo the mode switch from the fundamental mode to
the first overtone or vice versa, therefore the temporal dependence
of $\Pi$ as a function of $\tev$ undergoes the jump.
The horizontal branch evolution time is several orders of magnitude
greater than the pulsation period, so that without loss of accuracy
we can assume that the mode switch occurs instantaneously.
The evolutionary time corresponding to the mode switch was evaluated
in the present study as an average value of evolutionary times of two
adjacent hydrodynamical models pulsating in different modes.
For these models we determined the periods of the fundamental mode and
the first overtone using the discrete Fourier transform of the kinetic
energy.
Thus, in the point of the mode switch the periods of both modes are known.

\begin{figure}
\centerline{\includegraphics[width=15cm]{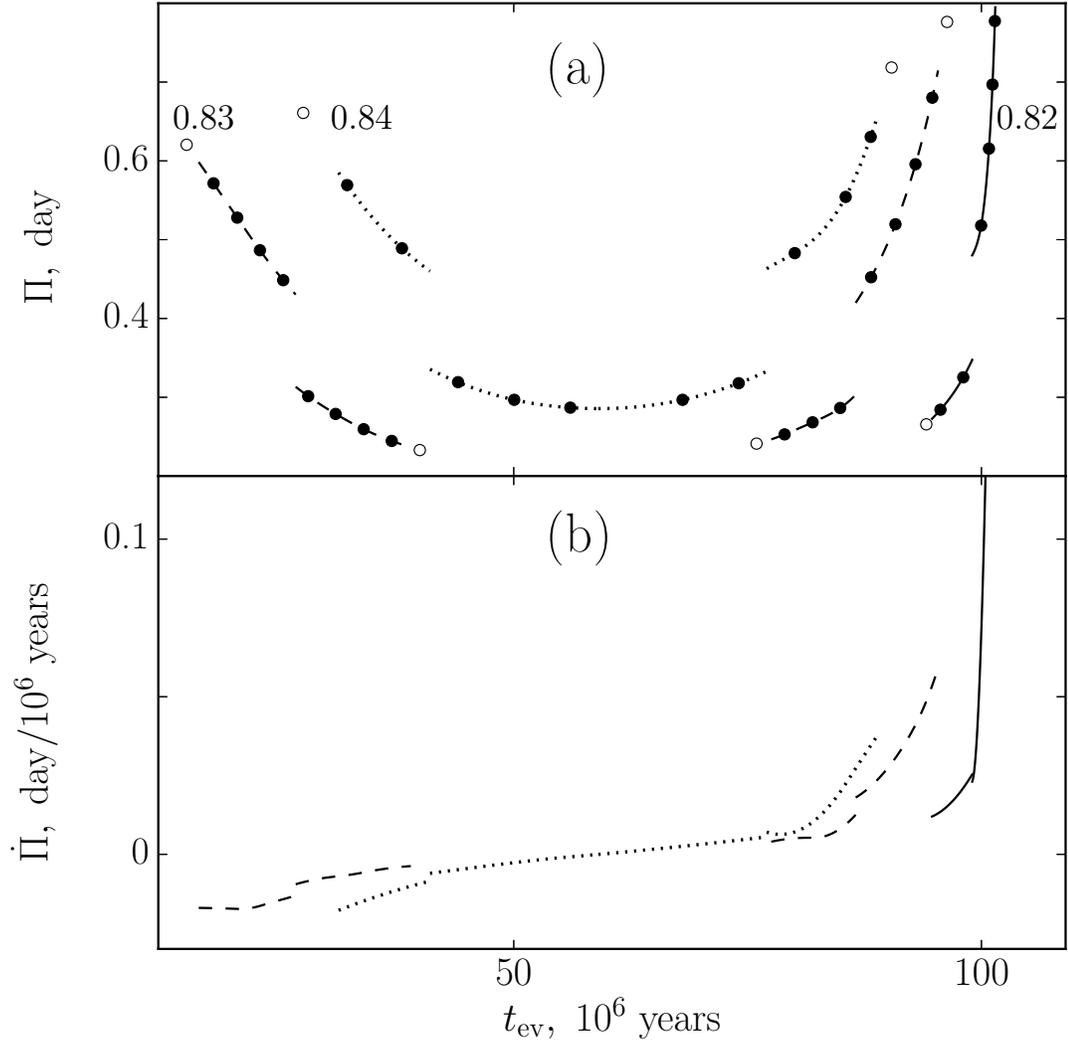}}
\caption{The pulsation period (a) and the period change rate (b)
as a function of evolutionary time $\tev$ for evolutionary sequences
$0.82$ (solid lines), $0.83$ (dashed lines) and $0.84M_\odot$ (dotted lines).
Periods of the models unstable against radial oscillations ($\eta>0$)
are shown by filled circles and the periods with decaying oscillations
($\eta<0$) are shown by open circles.}
\label{fig2}
\end{figure}

For the interval of evolutionary time $\tev$ with oscillations in one mode
the temporal dependence of the pulsation period was approximated by the
cubic interpolation spline.
Results of approximation are illustrated in Fig.~\ref{fig2}a for evolutionary
sequences $\mzams=0.82$, 0.83 and $0.84M_\odot$, where
for the sake of convenience the evolutionary time $\tev$ is set to zero
at the ZAHB point.
The shape of the plots depends on the location of the ZAHB point
and the minimum stellar radius (i.e. where the track turns redward)
on the HRD.
For $\mzams=0.84$ the stellar radius reaches its minimum within the
instability strip, so that abrupt changes in period are due to mode switch.
For $\mzams=0.83M_\odot$ the stellar radius reaches its minimum beyond the
blue edge of the instability strip at effective temperature $T_\mathrm{eff}=7930$~K,
so that stellar pulsations cease
for $39\times 10^6~\mathrm{yr} \le\tev\le 78\times 10^6~\mathrm{yr}$.
For $\mzams=0.82M_\odot$ the ZAHB point is beyond the blue edge of the
instability strip and radial oscillations arise during the stage of decreasing
effective temperature.

\section{the rate of period change in RR~Lyr type stars}

The rate of period change as a function of $\tev$ for
evolutionary sequences $\mzams=0.82$, 0.83 and $0.84M_\odot$ are shown
in Fig.~\ref{fig2}b where $\dot\Pi$ is given in units of $\mathrm{day}/10^6\mathrm{yr}$.
As can be seen, the rate of period change of the major part of RR~Lyr stars
ranges within $-0.02\le\dot\Pi\le 0.05\ \mathrm{day}/10^6\mathrm{yr}$ and
only the insignificant group of variables of the evolutionary sequence
$\mzams=0.82M_\odot$ near the red edge of the instability strip have the
period change rate greater than $0.1\ \mathrm{day}/10^6\mathrm{yr}$.

For understanding the evolution of RR~Lyr type variables in the globular cluster
the most interesting are results of comparison of the observed mean period change
rate with theoretical estimate of this quantity.
To this end, in the present study we employed the method of population synthesis.
We assumed that within the considered range of initial masses $\mzams$
the initial mass function is described by the uniform distribution.
The use of the uniform mass distribution is justified by the small width of
the initial mass interval: $\Delta\mzams/\mzams = 0.048$.

\begin{figure}
\centerline{\includegraphics[width=15cm]{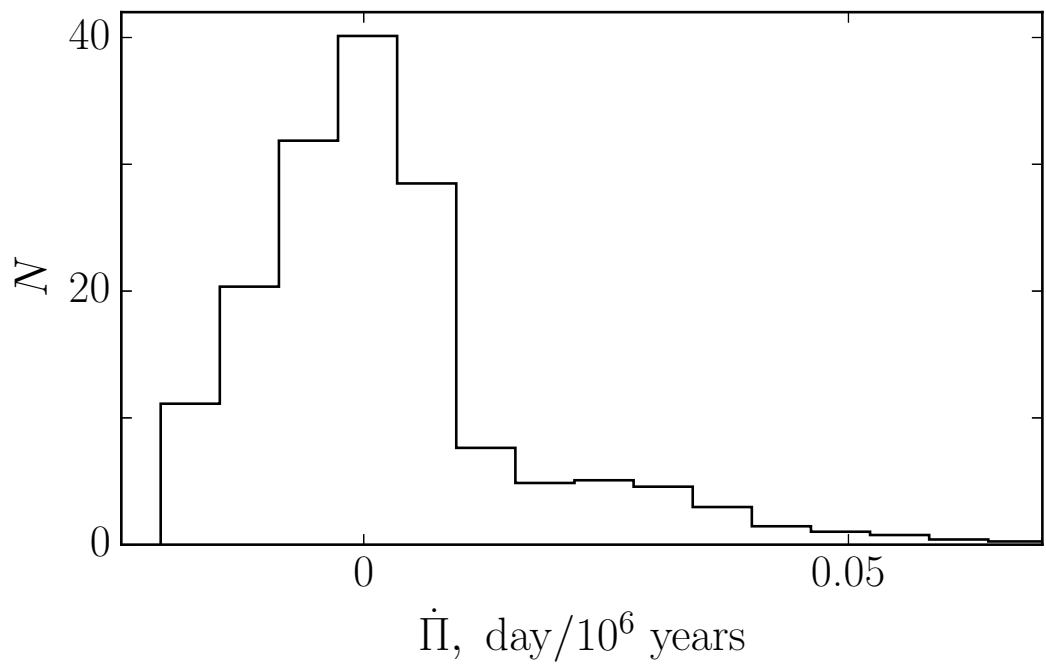}}
\caption{Normalized distribution of the number of RR~Lyr stars as a function
of period change rate.}
\label{fig3}
\end{figure}

Calculations were carried out with the Monte Carlo method, that is each random number
within $[0,1]$ was considered as a fraction of the lifetime on the helium burning
stage.
Once the initial mass of the star and its evolutionary time correspond to the location
within the instability strip, the program evaluates $\dot\Pi$ using
two--dimensional interpolation with respect to $\mzams$ and $\tev$.
The normalized distribution of the period change rates is shown in Fig.~\ref{fig3}
and the mean perion change rate of this distribution is
$\langle\dot\Pi\rangle=6.0\times 10^{-3}~\mathrm{day}/10^6\mathrm{yr}$.

\section{conclusion}

Results of stellar evolution and nonlinear pulsation calculations are in
a good agreement with observational estimate of the mean period change rate
$\langle\dot\Pi\rangle\approx 0.01~\mathrm{day}/10^6\mathrm{yr}$
obtained for RR~Lyr type stars of the globular cluster M~3 (Jurcsik et al. 2012).
In particular, such an agreement corroborates the theory of the core helium flash
which preceeds the horizontal branch evolutionary stage.
At the same time it has to be noted that the role of some parameters of
evolutionary computations in theoretical estimates of $\langle\dot\Pi\rangle$
remains unclear.
At first, we should bear in mind the initial fractional abundance of helium $Y$
and the parameter $\eta_\mathrm{R}$ in the Reimers formula (\ref{freim})
because both these quantities affect location of ZAHB on the HRD
and thereby change contribution of stars with different values of $\dot\Pi$
into the general distribution of the period change rate.

In the present study the method of population synthesis is based on
consistent stellar evolution and nonlinear stellar oulsation calculations,
whereas $\dot\Pi$ was evaluated using interpolation with respect to
two variables: the initial mass $\mzams$ and the evolutionary time $\tev$.
Therefore, the accuracy of the distribution can be improved by using
the denser grid of evolutinary tracks.
Our approach to population synthesis differs from traditional methods
(Silva Aguirre et al. 2008; Valcarce and Catelan 2008; Denissenkov et al. 2017)
where pulsational characteristics of stars are obtained from approximate relations
and therefore might be affected by significant uncertainties.
At the same time we should note that our approach needs much more
calculations beause besides the stellar evolutionary tracks one needs to
compute a great deal of hydrodynamic models.

\newpage
\section*{references}

\begin{enumerate}

\item F. Caputo, Astron. Astrohys. Rev. \textbf{9}, 33 (1998).

\item P.A. Denissenkov, D.A. VandenBerg, G. Kopacki, and J.W. Ferguson, Astrophys. J. \textbf{849}, 159 (2017).

\item A. Dotter, A. Sarajedini, and J. Anderson, Astrophys. J. \textbf{738}, 74 (2011).

\item Yu.A. Fadeyev, Pis'ma Astron. Zh. \textbf{39}, 829 (2013)
      [Astron. Lett. \textbf{39}, 746 (2013)].

\item Yu.A. Fadeyev, Pis'ma Astron. Zh. \textbf{40}, 341 (2014)
      [Yu.A. Fadeyev, Astron. Lett. \textbf{40}, 301 (2014)].

\item Yu.A. Fadeyev, MNRAS \textbf{449}, 1011 (2015).

\item J. Jurcsik, G. Hajdu, B. Szeidl, K. Ol\'ah, J. Kelemen, \'A. S\'odor, A. Saha, P. Mallick, and J. Claver,
      MNRAS \textbf{419}, 2173 (2012).

\item R.A. Koopmann, Y.-W. Lee, P. Demarque, and J.M. Howard, Astrophys. J. \textbf{423}, 380 (1994).

\item R. Kuhfu\ss, Astron. Astrophys. \textbf{160}, 116 (1986).

\item Y.W. Lee, Astrophys. J. \text{367}, 524 (1991).

\item B. Paxton, L. Bildsten, A. Dotter, F. Herwig, P. Lesaffre and F. Timmes,
      Astropys. J. Suppl. Ser. \textbf{192}, 3 (2011).

\item B. Paxton,  M. Cantiello,  P. Arras,     L. Bildsten,
      E.F. Brown, A. Dotter,     C. Mankovich, M.H. Montgomery, et al.,
      Astropys. J. Suppl. Ser. \textbf{208}, 4 (2013).

\item B. Paxton,   P. Marchant,  J. Schwab, E.B. Bauer,
      L. Bildsten, M. Cantiello, L. Dessart, R.  Farmer, et al.,
      Astropys. J. Suppl. Ser. \textbf{220}, 15 (2015).

\item B. Paxton,    J. Schwab,  E.B. Bauer, L. Bildsten, 
      S. Blinnikov, P. Duffell, R. Farmer,  J.A. Goldberg, et al.,
      Astropys. J. Suppl. Ser. \textbf{234}, 34 (2018).

\item D. Reimers, \textit{Problems in Stellar Atmospheres and Envelopes},
      Ed. by B. Baschek, W.H. Kegel, and G. Traving
      (Springer, New York, 1975), p. 229.

\item M. Schwarzschild and R. H\"arm, Astrophys. J. \textbf{136}, 158 (1962).

\item V. Silva Aguirre, M. Catelan, A. Weiss, and A.A.R. Valcarce, 
      Astron. Astrophys. \textbf{489}, 1201 (2008).

\item V. Silva Aguirre, M. Catelan, A. Weiss, and A.A.R. Valcarce,  Astrophys. Space Sci. \textbf{328}, 123 (2010).

\item A.V. Sweigart and A. Renzini, Astron. Astrophys. \textbf{71}, 66 (1979).

\item A.A.R. Valcarce and M. Catelan, Astron. Astrophys. \textbf{487}, 185 (2008).

\item D.A. VandenBerg, P.A. Denissenkov, and M. Catelan, Astrophys. J. \textbf{827}, 2 (2016).

\end{enumerate}

\newpage
\clearpage

\newpage
\clearpage

\newpage
\clearpage

\end{document}